\documentclass[preprint2]{aastex}

\shorttitle{Photometry and Luminosity Functions of M92}
\shortauthors{Nathaniel E. Q. Paust and Brian Chaboyer}

\begin{document}

\title{BVI Photometry and the Luminosity Functions of the Globular Cluster M92}
\author{Nathaniel E. Q. Paust and Brian Chaboyer}
\affil{Department of Physics and Astronomy, Dartmouth College, 6127 Wilder Laboratory, Hanover, NH 03755}
\email{nathaniel.e.paust@dartmouth.edu}
\and \author{Ata Sarajedini}
\affil{Department of Astronomy, University of Florida, 211 Bryant Space Science Center, P.O. Box 112055, Gainesville, FL 32611}

\begin{abstract}
We present new BVI ground-based photometry and VI space-based photometry for the globular cluster M92 (NGC 6341) and examine luminosity functions in B, V, and I containing over 50,000 stars ranging from the tip of the red giant branch to several magnitudes below the main sequence turn off.  Once corrected for completeness, the observed luminosity functions agree very well with theoretical models and do not show stellar excesses in any region of the luminosity function.  Using reduced-$\chi^2$ fitting, the new M92 luminosity function is shown to be an excellent match  to the previously published luminosity function for M30.  These points combine to establish that the ``subgiant excess'' found in previously published luminosity functions of  Galactic globular clusters are due to deficiencies in the stellar models used at that time.  Using up to date stellar models results in good agreement between observations and theory.

Several statistical methods are presented to best determine the age of M92.  These methods prove to be insensitive to the exact choice of metallicity within the published range.  Using [Fe/H]=$-2.17$ to match recent studies we find an age of $14.2 \pm 1.2$ Gyr for the cluster.

\end{abstract}


\section{Introduction}
Globular clusters are ideal locations to test stellar evolutionary models due to their single-age single-metallicity nature.  Previously, work has focussed in great depth on analysis of the color-magnitude diagram (CMD) of clusters, for example in \citet{rfp}.  While studies of the CMD can reveal a great deal about stellar evolution, the luminosity function (LF) of the cluster is especially powerful for determining the timescales of stellar evolution.  Below the main-sequence turnoff (MSTO), the LF is primarily a reflection of the initial mass function modulated by dynamical mass segregation effects at the lowest masses.  However, above the MSTO, the LF reveals the progress of the hydrogen burning shell through the star and can even give hints as to the internal structure of the star.  Indeed, the enhancement in the LF known as the LF 'bump' marks the hydrogen-burning shell's transition from the region previously mixed by convection into unmixed stellar material \citep{iben}.

Only with the advent of large-format CCD cameras has it become possible to obtain precise photometry of a large numbers of stars in all phases of stellar evolution.  Previous studies have been largely limited to either low precision photographic measurements or small spatial coverage and limited samples of stars.  With detailed LFs, unexpected results have appeared.  \citet{bolte} found an excess of stars on the subgiant branch (SGB) for the low metallicity, [Fe/H]=$-2.12$ \citep{harris}, cluster M30 (NGC 7099).  The subgiant excess in M30 was further confirmed by \citet{sb2} using a higher-quality data set.  This SGB excess is the expected result if weakly interacting massive particle (WIMP) energy transport is important in stars \citep{fs}.  Recent LFs of the more metal rich clusters M5 \citep{sb} and  M3 \citep{rood} with [Fe/H]=$-1.27$ and $-1.57$ respectively do not show the SGB excess, suggesting that it may only be a characteristic of metal-poor clusters.  However, the models used in the comparison do not include diffusion, which is a standard part of modern stellar evolution codes and the statistical basis for the SGB excess is not explicitly defined.

The luminosity function of M92 (NGC 6341) is an ideal test for stellar modeling codes.  Its metallicity, [Fe/H]=$-2.27$ \citep{harris}, places it in the same abundance range as M30, which was earlier found to have a SGB excess.  It is a large cluster, making it easy to measure significant numbers of stars.  Its location, far from the galactic plane ($b=34.86$ degrees \citep{harris}), minimizes the significance of field-star contamination.  Finally, it is a fairly well studied cluster with accurately determined distance modulus and metallicity which simplifies comparisons of the observed LF to models.  Finally, previous LF studies of the cluster such as \citet{piottock} and \citet{lee} have not looked at the LF along the RGB in order to examine the previously found excess present in M30, instead they have concentrated on the lower MS.  The \citet{lee} study further supports using M92 as a test for stellar modeling codes since examination of the mass function in that work suggests that the cluster has not been strongly affected by tidal shocks resulting in a pure sample of cluster stars.

In the context of the above points, we herein present our analysis of a comprehensive luminosity function study of M92 using ground-based and space-based observations.  The next section describes these observations, while Section 3 discusses the reduction of the data.  Sections 4 and 5 present the color-magnitude diagrams and luminosity functions.  Section 6 details the stellar evolution model and discusses details in fitting the theoretical models to the observed LFs.  Section 7 compares the new M92 LF to previous LFs of M92 and M30 and Section 8 presents conclusions about the fits and the general state of LF modeling.

\section{Observations}
The ground-based images were collected from 21-27 April 2003 using the Hiltner 2.4m telescope at MDM Observatory on Kitt Peak.  All frames were taken using the Echelle 2048x2048 camera with a pixel scale of 0.28 arsec/pixel and a total field of view of 9.56 arcminutes.  To cover a significant fraction of the cluster area, short and long exposures were tiled over a 27x27 arcminute area.  For the short images, exposure times were chosen to ensure that stars at the tip of the red giant branch (RGB) were not saturated,  typically resulting in 5-15 second exposures.  Two images in each filter were taken at each grid position for the short images.  The exposure lengths for the longer exposures were in the range of 60-120 seconds depending on the filter and 3 exposures were taken in each filter at each grid position.  All of the images were observed under photometric conditions with average seeing of approximately 0.8 arcseconds.  

The Hubble Space Telescope images were taken from the STSCI archives and were collected on 27 August, 2002, using the Advanced Camera for Surveys.  The Hubble Space Telescope images were obtained with the Advanced Camera for Surveys on 27 August 2002 as part of HST program GO-9453. The exposures cover the central 3-arcminutes of the cluster in the F606W and F814W filters, with exposure lengths of 0.5, 5, and 90 seconds and 0.5, 6, and 100 seconds respectively in the two filters.  Because these observations were taken shortly after ACS was installed, the charge transfer efficiency corrections are negligible \citep{brown}.  

\section{Data Reduction}
\subsection{Image Processing}
The ground-based images of M92 were processed using standard IRAF techniques utilizing evening and dawn twilight flats.  The archival Hubble images were processed using the STSCI on-the-fly-reprocessing system from the Multi-mission Archive at Space Telescope (MAST).  The drizzled images, multiplied by the image exposure time, were utilized for the photometry.

\subsection{Photometry}
\subsubsection{The ground-based sample}
Profile fitting photometry was performed on the 179 image frames using the DAOPHOT and ALLSTAR programs developed by Peter Stetson \citep{stet87,stet94}.  In each frame, approximately 100 bright uncrowded stars were chosen to determine the point-spread function (PSF) and its variation about the frame.  While the PSF stars were specifically chosen to be outside the most crowded regions of each frame, they generally did sample a large fraction of the frame and therefore map the PSF variations well.  Several scripts were used to automate two passes through DAOPHOT's FIND routine and ALLSTAR to generate a photometry list for each image.

Photometry from the individual frames was filtered to remove any detection with a measured error greater than 0.1 mag.  Aperture corrections were calculated using the
brightest uncrowded stars on each frame. These were used to search for a spatial dependence of the aperture correction, but none was found.  DAOMASTER and DAOMATCH were then used to combine the individual photometry files into one master file for each filter, requiring that a star be detected in at least two frames in each filter.  These master files were filtered to contain only stars whose frame-to-frame magnitude variation was less than 0.1 magnitude.  The B, V, and I master files were then matched using DAOMASTER and DAOMATCH requiring a star to be detected in all three filters to be included in the final catalog. 

\subsubsection{The HST sample}

The space-based images were processed using the same methods as the ground based data, although the process was simplified by the fact that there were only 6 images.    While the same criteria for rejecting stars based on photometric errors was used, the matching criterion was relaxed to require that stars only be detected once in each filter.

\subsection{Calibration}
\begin{figure}
\plotone{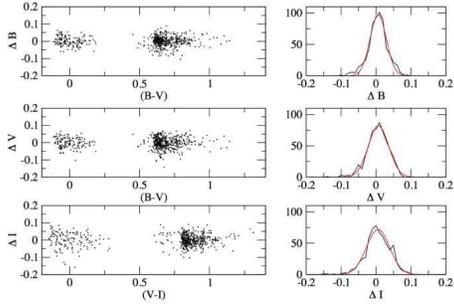}
\caption{Residual distributions for the standard stars. The second column of plots shows the error distribution displayed with fitted gaussians with sigma 0.023, 0.028, and 0.033 for B, V, and I.  These sigmas are as expected for simple error propagation given the photometric errors.}
\label{fig:standardresid}
\end{figure}

The instrumental magnitudes were brought onto the standard system using P. Stetson's photometric standards for M92 \citep{stet00}.  Using stars brighter than 17th magnitude in V, we were able to find over 600 stars with B, V, and I magnitudes in common between the Stetson standards and our ground-based data.  The best photometric solution to bring the data to the Stetson system was found to depend on color only to first-order.  The transformation equations were determined to be
$$B=b+2.685+0.0537 \ (B-V)$$
$$V=v+1.530-0.0413 \ (B-V)$$
$$I=i+0.223+0.0120 \ (V-I)$$
The residuals from the fit between the observed stars and the Stetson standards are shown in Figure \ref{fig:standardresid}.  In all cases, the distribution of the residuals is exactly the distribution expected from the photometric errors.

\begin{figure}
\plotone{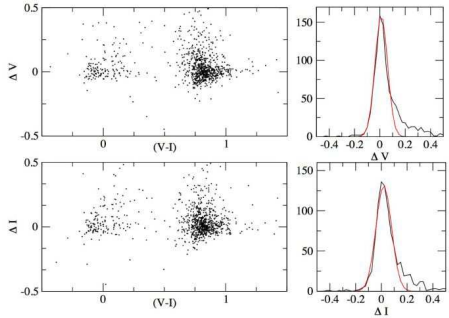}
\caption{Residual distributions for the standardized HST sample.  The residual distribution is shown overplotted with gaussians with standard deviations of 0.05 in V and 0.06 in I.}
\label{fig:hstresid}
\end{figure}

The HST data posed an interesting problem in calibration because of the small field of view in the crowded core of the cluster.  Due to confusion between stars in the core of the cluster, there were no Stetson standard stars in the field of view of the HST images.  It was possible, however,  to calibrate the HST photometry using our ground-based observations as secondary standards.  Using the same method as in the calibration of the ground-based data, approximately 400 stars in common were compared.    These stars all lie in the outer 0.5 arcminutes of the HST images.  In order to simplify the interpretation of the luminosity functions, which are based on BVI magnitudes, the instrumental magnitudes from the Hubble data were transformed directly into V and I using the following relationships
$$V=606_{inst}+0.4881+0.0695 \ (V-I)+0.1510 \ (V-I)^2$$
$$I=814_{inst}-0.3850-0.0521 \ (V-I)+0.0306 \ (V-I)^2$$
Figure \ref{fig:hstresid} shows the residuals between the HST photometry and the ground based photometry.  The distribution matches expectations based on the photometric errors in the two data sets.

\section{The Color-Magnitude Diagrams}
\begin{figure}
\plottwo{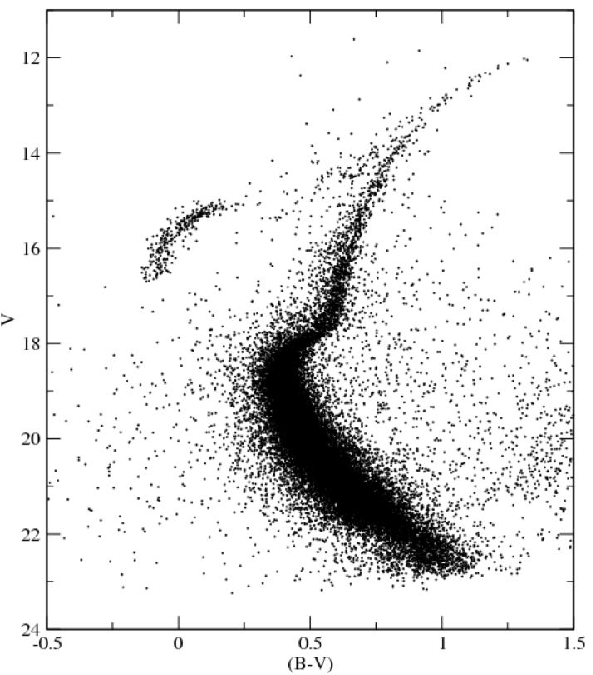}{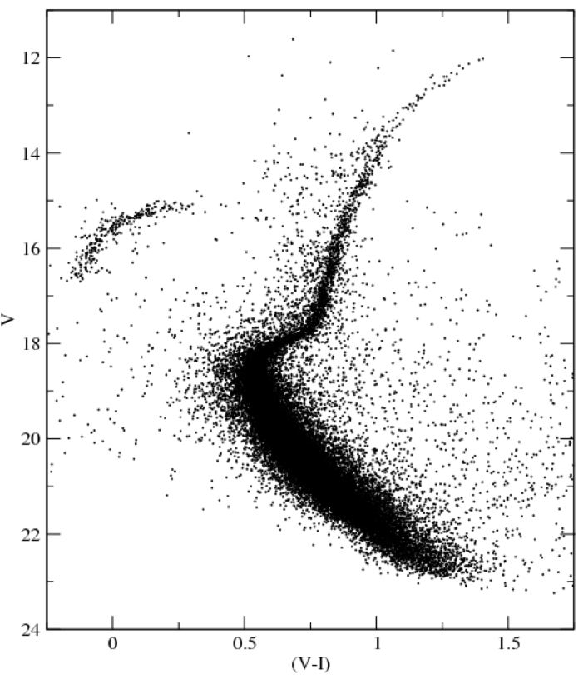}
\caption{The V -- $(B-V)$ and V -- $(V-I)$ color-magnitude diagrams for the ground-based sample.}
\label{fig:groundcmd}
\end{figure}

\begin{figure}
\plotone{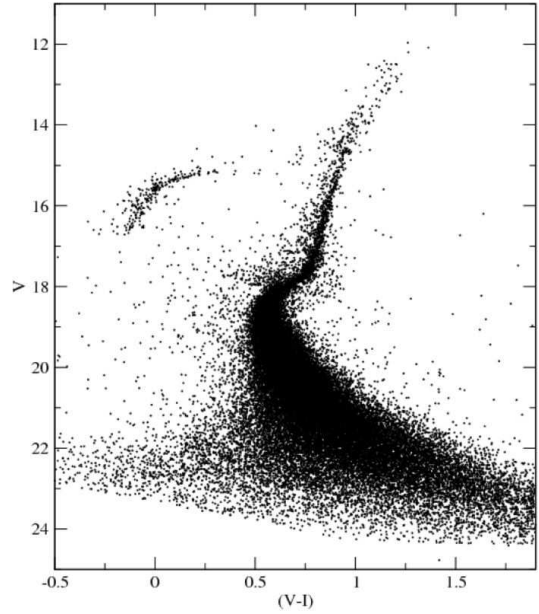}
\caption{The V -- $(V-I)$ color-magnitude diagram for the HST sample.  This CMD covers an area of approximately 3x3 arcminutes and contains over 41,000 stars.}
\label{fig:hstcmd}
\end{figure}

In total, 34,242 stars  in the ground-based sample and 41,205 stars in the HST sample were measured.  The HST sample is larger due to the decreased confusion in the core which allowed a much larger number of MS stars to be detected.  V versus $(B-V)$ and $(V-I)$ color-magnitude diagrams from the terrestrial sample are shown in Figure \ref{fig:groundcmd}.  The V versus $(V-I)$ color-magnitude diagram from the HST sample is shown in Figure \ref{fig:hstcmd}. 

\section{The Luminosity Functions}
\subsection{Completeness}
The critical task in generating a luminosity function is determining the completeness of the photometric sample.  Generally, the completeness  is a function of both position relative to the cluster, due to the confusion in crowded regions, and magnitude.  An accurate determination of the completeness is found through extensive use of artificial star tests.  These tests were performed by adding artificial stars to each image and then remeasuring photometry for the complete field to determine what fraction of the added artificial stars are recovered.  Errors in the eventual LFs are governed by Poisson statistics, so large numbers of input and recovered artificial stars are needed.  However, adding a large number of stars to any particular image will change the crowding and therefore the completeness.  In order to balance this effect,  many runs through the artificial star routine are used with relatively small numbers of stars added in each run.  

\begin{figure}
\plotone{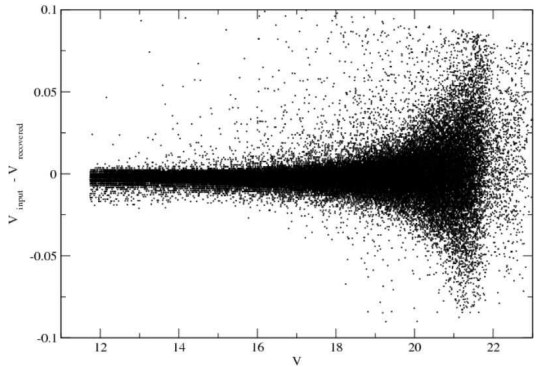}
\caption{The difference between input and recovered artificial star magnitudes for the 59,000 recovered artificial stars as a function of V.  The average bias is insignificant at -0.0005 magnitudes in V and similar amounts in B and I.}
\label{fig:inoutbias}
\end{figure}

The general method used for the artificial star tests was as follows: a master list of artificial stars with random x and y pixel positions covering the entire area imaged was created.  The magnitude for the artificial stars was determined by generating random numbers covering the range of instrumental B magnitudes found during the photometry.  The assigned B magnitudes were then used with the CMD ridge lines to assign V and I magnitudes to each artificial star.  The result is a list of artificial stars distributed randomly across the imaged area with magnitudes and colors that match the observed cluster stars.   Each star from the master list was placed at the appropriate position with the appropriate instrumental magnitude in each frame that it could have appeared.  The image frames with added artificial stars were then put  back through the same DAOPHOT and ALLSTAR pipeline used for the initial photometry including the same error clipping and required number of detections during matching.  Additionally, the input and recovered magnitudes for the artificial stars were required to be within 0.1 mag in order to remove the possibility of real stars or blends being confused with the artificial stars.  Figure \ref{fig:inoutbias} shows the difference between the input and recovered artificial star magnitudes as a function of V magnitude.  To investigate any magnitude bias in the recovered artificial stars, the stars were divided into bins two  magnitudes wide and the average difference between the input and recovered magnitude was determined.  For all bins brighter than 22th magnitude, the bias had an absolute value of $0.002$ magnitudes or less.  The final bin had bias of $0.012$. However, the last bin is beyond the completeness limit and has an order of magnitude less stars than any of the other bins.  Based on this analysis, there is no significant bias in the artificial star magnitudes.  In all, 24 sets of artificial stars, taking approximately 36 cpu-hours per set, were added to the images for a total of over 91,000 input stars and 59,000 recovered stars.

\begin{figure}
\plotone{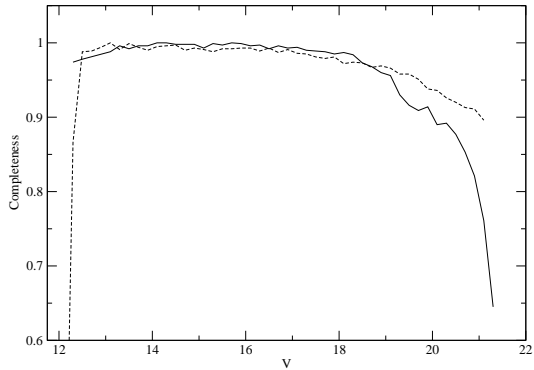}
\caption{The completeness of the ground-based (solid line) and space-based (dashed line) data as a function of V magnitude.  The completeness limits quoted for the LFs are slightly different than may be expected from these lines since they are based on completeness as sampled by the positions of the cluster stars rather the random positions of the artificial stars.}
\label{fig:completeness}
\end{figure}

Due to the relatively small number of images in the space-based data, it was possible to use a larger number of artificial star runs in a reasonable amount of time.  Using the same procedure, over 267,000 artificial stars were added and over 203,000 were recovered from the images after processing 90 sets of artificial stars.  The completeness of the datasets as a function of magnitude is shown in Figure \ref{fig:completeness}.  The lower completeness at the bright end is due to bright stars saturating in some of the images.

\subsection{Computing the Luminosity Function}

There are two main steps in transforming the photometric data into the luminosity function.  First, field stars, horizontal branch (HB) stars, and asymptotic giant branch (AGB) stars must be removed to produce a pure sample of stars on the MS and RGB.  The field stars, HB stars, and AGB stars were removed by clipping all stars more than 3-$\sigma$ in color from the fiducial reference line.  To find the fiducial line, a box was run across the CMD in color at different magnitudes.  The set of points in color that maximize the population in the box as a function of magnitude define the fiducial line.  This method will not remove any field stars that lay along the MS and RGB, but the density of field stars is low enough that the resulting LF is not significantly affected.  This is confirmed by the Besancon stellar population synthesis models \citep{robin} which predict approximately ten field stars along the upper half of the RGB.  The photometric clipping method is ambiguous at the top of the RGB where the AGB and RGB nearly merge.  In this region, extraneous AGB stars were removed manually.

Completeness corrections were accomplished by weighting each star individually.  A grid of completeness as a function of radius from the cluster center and magnitude was constructed and then used to assign a weight of $\hbox{completeness}(m,r)^{-1}$ to each star.  The LF was then constructed by summing the weights of stars in each magnitude bin.  The completeness limit was defined as the magnitude where the average weight per star in the bin was equal to 2, roughly the point where the data is 50\% complete.  The completeness in the V band as a function of magnitude is shown in Figure 6.  The error bars on the luminosity function are Poisson errors given by $$\sigma(N_i)=N_i \sqrt{1/n_i+1/a_i+1/b_i}$$ where $N_i$ is the summed weights of stars in the magnitude bin, $n_i$ is the number of stars in the magnitude  bin, $a_i$ is the number of recovered artificial stars in the magnitude bin, and $b_i$ is the number of input artificial stars in the magnitude bin.  The number of stars in each bin is the dominant source of uncertainty rather than the number of input or recovered artificial stars.

\begin{figure}
\plotone{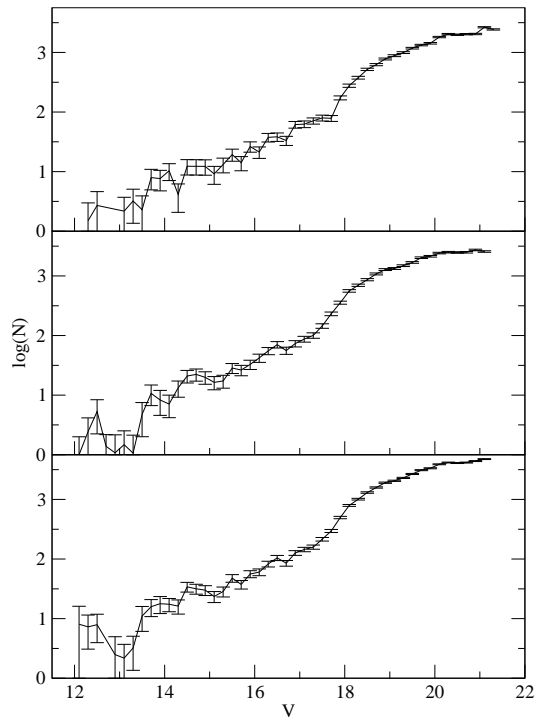}
\caption{From top to bottom:  the ground-based V LF, the HST V LF, and the combined V LF.  In all three LFs, the error bars are the 1-$\sigma$ Poisson errors. The LFs in B and I look very similar and are of equal quality.}
\label{fig:threelfs}
\end{figure}

This weighting method revealed one shortcoming of the ground-based data, namely the fact that, due to the high stellar density, no stars fainter than the SGB were detected in the central two arcminutes of the cluster.  This introduces a skew into the LF since the area being sampled at each magnitude is different.  This could be corrected by defining a cluster profile based on brighter, more complete, stars and then correcting the numbers of fainter, less complete, stars.  This method has been used in previous papers in the literature.  However, any correction would assume that the cluster has the same profile for high and low mass stars and would completely disregard any mass segregation in the cluster, as was found for M3 by \citet{rood}.  This method could also cause problems since errors in the numbers of bright stars, with poor error statistics, would be propagated through the entire LF.  Also, because the size of the central hole depends very strongly on magnitude, extreme care would be required to avoid artificially generating an excess or deficiency at points in the LF.  To simplify the analysis of the LF, the central area of the ground based data, containing approximately 11,000 stars, was removed.  The resulting completeness limits are B=22.1, V=21.3, and I=20.5.  The V luminosity functions from the ground, HST, and combined data sets are shown in Figure \ref{fig:threelfs} and are representative of the LFs in the other bands.  The ground based LFs are listed in Table 1 in the appendix.

The HST sample displayed reasonable completeness across the entire field, unlike the ground based sample, so the entire catalog was used to create the luminosity functions using the procedure outlined above.  Due to the relatively small field of view of the ACS, very few stars near the tip of the RGB were detected resulting in a very noisy LF.  The completeness limits are V=21.1 and I=20.1, limited largely by the short exposure times.  The LFs from the HST sample are listed in Table 2 in the appendix.

Combining the HST data from the cluster's central region with the ground-based photometry of the outer region, it was possible to create a composite LF covering a full 27 arcminute square area of the cluster.  The composite LF contains data for over 54,400 weighted stars.  The combined LFs are listed in Table 3 in the appendix.  The completeness limits for this sample are taken to be the completeness limits of the HST sample.

\section{Theoretical LF Models and Fitting}

To produce the theoretical LFs for comparison to the observed data, stellar models were created using the Dartmouth Stellar Evolution Code (DSEP) \citep{chaboyer}.  These models used the \citet{vandenberg} color transformations.  A catalog of theoretical luminosity functions was created with ages stepped by 0.1 Gyr from 11.6  Gyr to 16.5 Gyr.  The models use scaled solar compositions, while globular clusters are typically enhanced in their $\alpha$-element (O, Mg, Si, S, and Ca) abundances.  As noted by \cite{chi91} and \cite{cha92},  scaled solar composition models are nearly identical to $\alpha$-element enhanced models, provided one modifies the relationship between [Fe/H] and the heavy element mass fraction $Z$.   The modification assumed  $[\alpha/\mathrm{Fe}] = +0.40$.   Metallicities of [Fe/H]$=-2.11$, $-2.14$, $-2.17$, $-2.20$, $-2.23$, $-2.27$, and $-2.31$ were chosen to sample the published range of $-2.24 \pm .08$ \citep{zw}, $-2.27$ \citep{harris}, and $-2.38 \pm 0.07$ \citep{kraft}.  The majority of the comparisons between the observed and theoretical LFs were completed using the [Fe/H]$=-2.17$ models since they are the closest metallicity match to the recent \citet{kraft} measurements after being modified for the scaled-solar assumption.  All of the statistical tests proved to be insensitive to metallicity in the studied range.

One of main parameters in fitting the theoretical LFs to the observed ones is distance modulus.  To ensure that this represents the absolute distance modulus, the observed LFs must be corrected for extinction.   A standard extinction-color excess relation was used, $A(V)=3.1 \ E(B-V)$, with the published value of the color excess being $0.02$ for M92 \citep{harris}.  Extinction in the I band was calculated assuming $E(V-I)=1.25 \ E(B-V)$ \citep{dwc}.  All extinctions were computed on the B,V, and I magnitudes.  \citet{sirianni} suggest that reddening and extinction should be considered in the native 606W and 814W filters for the space-based observations due to the slight differences with the standard V and I filters.  Due to the low reddening to M92, the difference in methodology corresponds to a difference of 0.008 magnitudes in the V filter and 0.001 magnitudes in the I filter.

\begin{figure}
\plotone{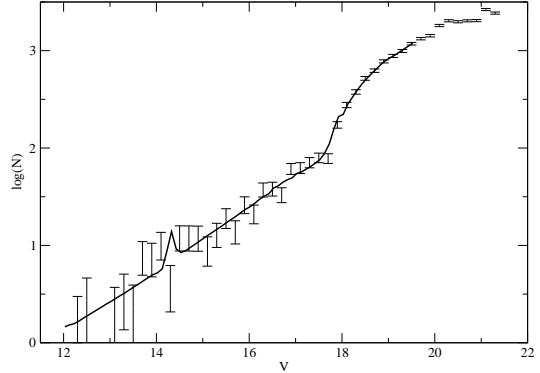}
\caption{The theoretical 13.9 Gyr [Fe/H]$=-2.17$ LF, the solid line, plotted with the observed V LF from the ground-based sample, the error bars.  The reduced-$\chi^2$ is $0.96$.  The B and I fits have equal quality.}
\label{fig:groundVfit}
\end{figure}

Three separate statistical methods were used to determine the best match between the observed and theoretical LFs.  First, a reduced-$\chi^2$ fitting method based on \citet{br} was used to find the best match between the theoretical and observed LFs in all three filters simulatinously.  During the reduced-$\chi^2$ minimization, the absolute distance modulus was allowed to vary while the normalization was set by the total number of stars brighter than the completeness limit in the LF.  Regardless of the theoretical model used,  the reduced $\chi^2$ values were similar with values around 0.9 with a wide range of distance moduli ($\pm 0.3$ magnitudes) and ages ($\pm 1.2$ Gyr) producing seemingly reasonable fits.  The reduced-$\chi^2$ fits covered a range from the tip of the RGB to one magnitude below the MSTO, to $V=18.9$.  This covers a mass range from 0.73 to 0.77 $M_\odot$ and therefore tests the relative numbers of stars on the RGB and MS rather than being sensitive to the cluster IMF.  Standard probability tables for the reduced-$\chi^2$ statistic assume Gaussian errors while the errors in the LF come from Poisson counting.  As a result, we employ a Monte Carlo simulation to estimate the statistical significance of our LF comparisons.  In the simulation, $10^6$ realizations of the M92 LF were generated using the quoted errors.  The reduced-$\chi^2$ values between the original LF and the realizations were then computed to generate a probability table.  Reduced-$\chi^2$ values of 0.9 or less are found $75\%$ of the time showing a good match between the observed LF and the theoretical LFs.   A typical best match between the theoretical and observed ground-based LFs is shown in Figure \ref{fig:groundVfit}.

\subsection{The K-S Test}
In order to get a better constraint on the data,  the observed and theoretical data were compared using the Kolmogorov-Smirnov (K-S) test.  The K-S test maps the maximum difference between the theoretical cumulative LF and the observed cumulative LF and is therefore somewhat immune from the potential blending of stars between bins by photometric errors in the conventional LF.

\begin{figure}
\plotone{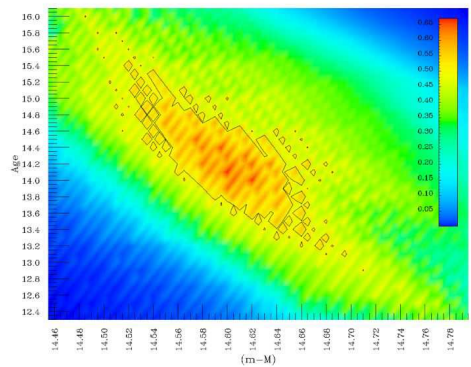}
\caption{The K-S test results for  [Fe/H]$=-2.17$ models.  The color scale gives the confidence as a percentage from the K-S test, the contours are drawn at the 50\% level.}
\label{fig:ksresult}
\end{figure}

To complete the K-S test, a subsection of the luminosity functions from approximately 0.5 mag fainter than the tip down to the MSTO was defined.  The cumulative LF from this region is then compared to the cumulative theoretical LF while stepping through  the absolute distance modulus. Ages from 11.6 Gyr to 16.5 Gyr were examined in 100 Myr increments.  The K-S probability statistic was then used to find the best match.  The error in age and absolute distance modulus was assumed to be given by the error ellipse traced by the $50\%$ contour line.  There was no significant difference between the different metallicity models so comparisons have been made using the model closest to the \citet{kraft} metallicity, the [Fe/H]$=-2.17$ model.  This results in good agreement between the observed LFs and the model giving an age of $14.2 \pm 1.2$ Gyr.  This result is shown in Figure \ref{fig:ksresult}.

\section{Comparisons}
\subsection{Previous M92 Luminosity Functions}
\begin{figure}
\plotone{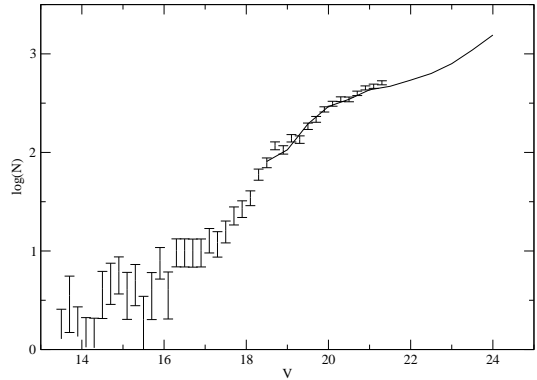}
\caption{The \citet{lee} LF and our LF in the V band.  Our LF has 1-sigma error bars while the Lee LF is plotted without error bars for clarity.}
\label{fig:leefit}
\end{figure}

\citet{lee} published a luminosity function for stars fainter than than the MSTO, which serves as a good check for the luminosity function presented in this paper on the low mass end.  The shape of the Lee et al. luminosity function agrees very well with our luminosity function over the range of V=18.5 to V=20.9 giving a reduced $\chi^2$ value of 0.52 once the two LFs are normalized.  The match between the LFs is shown in Figure \ref{fig:leefit}.

\subsection{The M30 Luminosity Function}
\begin{figure}
\plotone{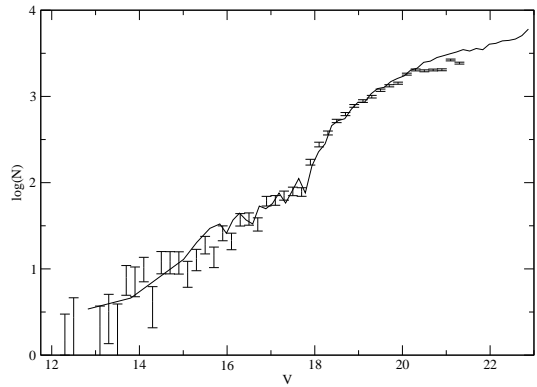}
\caption{The \citet{sb2} M30 LF and our LF with 1-$\sigma$ error bars in the V band.  For clarity, the error bars have been omitted from the M30 LF.}
\label{fig:m30m92}
\end{figure}

Previous work such as \citet{sb2} have shown that the LF of M30 has an excess of stars on the subgiant branch when compared with stellar models. It is not clear if this behavior is particular to M30 or if it is a general characteristic of low-metallicity globular clusters.  To compare the LFs from M30 and M92, the reduced-$\chi^2$ between the two cluster LFs ignoring the small metallicity difference was calculated.  To find the minimum reduced-$\chi^2$, the distance modulus and normalization of M30 were allowed to vary.  Assuming a distance modulus of 14.64 for M92 the best fits were found using a distance modulus of 14.92 for the V LF and a distance modulus of 14.82 for the I LF.  This matches well with the distance modulus of $14.87 \pm 0.12$ found by \citet{sb2} assuming a reddening of $E(V-I)=0.06$. The resulting reduced-$\chi^2$ values are 0.26 in V and 0.36 in I and suggest that the LFs for M30 and M92 are very similar. Given the excellent agreement between the theoretical LFs and the M92 LFs and given the extreme similarity between the M92 LFs and previous M30 LFs, there seems to be no problem with current stellar evolution models in contrast to previous results of \citet{sb2}. The match between the M30 and M92 V LFs are shown in Figure \ref{fig:m30m92} and is also representative of the quality of the fit in the I band.

Additionally, the \citet{sb2} LF was directly compared  to models with [Fe/H]=$-2.02$ and $-2.42$ to bracket the nominal metallicity.  Allowing for a slight uncertainty in the distance modulus and normalizing the theoretical LFs to match the observed LFs, reduced-$\chi^2$ values of 1.16 and 0.95 were found for the two models.  Using the generated probability table based on Poisson errors, we find that a reduced-$\chi^2$ value less than 0.95 is found $53\%$ of the time and a value of 1.16 is found $25\%$ of the time.  From this, we can conclude that the M30 LFs of \citet{sb2} are well fit by DSEP models.  

\begin{figure}
\plotone{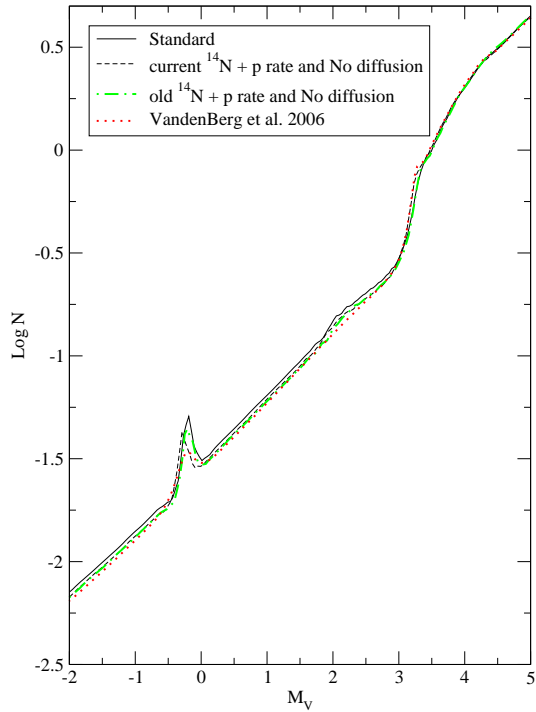}
\caption{14 Gyr DSEP LFs with and without current reaction rates and diffusion compared to \citet{vandenberg2006} LFs.  When normalized to the MS, the \citet{vandenberg2006} models have $9\%$ fewer stars along the RGB.  This decrease results in the excess of stars found in \citet{sb2}.}
\label{fig:vandsep}
\end{figure}

Adding further support to this argument is a direct comparison of the luminosity functions used in \citet{sb2} to current DSEP luminosity functions. Figure \ref{fig:vandsep} compares the standard DSEP model, including diffusion and the most recent reaction rate for  $^{14}N+P \rightarrow ^{15}O+\gamma$, the slowest rate in the CNO cycle, from \citet{formicola} to models with the old and new rates without diffusion and \citet{vandenberg2006} models.  All of the LFs are at the same metallicity and are 14 Gyr old. The \citet{sb2} paper utilized earlier versions of the \citet{vandenberg2006} models.  From the Figure, it is apparent that the \citet{vandenberg2006} models are similar to the DSEP no-diffusion models.  Compared to the standard DSEP model, the \citet{vandenberg2006} models have $9\%$ fewer stars along the RGB.  The difference explains the deviation found between the observed LF and the theoretical LFs in \citet{sb2}.

This underscores a point raised in \citet{gza2005} that a variety of different model libraries should be utilized before assuming some unaccounted phenomenon is responsible for deviations between observed and theoretical LFs.  It also gives a great amount of confidence in stellar evolution models which use up to date physics, such as DSEP.  Furthermore, the comparison between DSEP and \citet{vandenberg2006} LFs and the successes and failures in matching observed luminosity functions of low-metallicity GCs supports the inclusion of gravitational settling and microscopic diffusion in stellar evolution models and the implied reduction in globular cluster ages of order 1 Gyr.

\subsection{The RGB Bump}
The bump is caused by an increase in the hydrogen content of the material fed into the hydrogen-burning shell in stars along the RGB.  This increase comes when the shell passes through the former base of the convection zone \citep{iben}.  While it is difficult to make a concrete comparison due to the large error bars in the LF along the RGB,  after examining the best matches it appears that M92's bump is approximately 0.4 mag fainter than the theoretical bump.  This agrees with previous results from \citet{fusipecci} which found a 0.4 mag difference for a sample of 13 clusters.  Two possible explanations for the difference exist: either the hydrogen-burning shell has moved more rapidly than expected or convection has  penetrated farther into the star than expected, perhaps through a mechanism such as convective overshoot.  Additionally, recent results \citep{bjork} suggest that the location of the bump is extremely dependent on metallicity and composition with 0.2 magnitude or more changes being possible within reasonable ranges of metallicity and $\alpha$-enhancement.

\section{Conclusions}
1.  This paper finds good agreement between theoretical LFs and observations in terms of relative stars counts along the MS, SGB, and RGB in the M92 LF.   While the LFs are noisy in the RGB region, the theoretical LF matches the observed LF within the errorbars.  The result is further strengthened by the use of two independant data sets: the ground based observations and the HST observations.  

2.  Contrary to previously published results, we find no ``subgiant excess'' in comparisons of the M92 LF to DSEP LFs, the M92 LF to the \citet{sb2} of M30, or the M30 LF directly to the DSEP LFs.  It appears that the ``subgiant excess'' found in previous works is due to an erroneous under-prediction of the models rather than a real excess of stars in the cluster.  Direct comparison of DSEP LFs to \citet{vandenberg2006} LFs reveal that gravitational settling and microscopic diffusion all combine to better fit observations.  Thus, observations of metal-poor GCs support the inclusion of these effects in stellar evolution models and their implied reduction in globular cluster ages.

3.  Utilizing two separate methods to compare the theoretical and observed LFs it is possible to constrain the age of the cluster to $14.2 \pm 1.2$ Gyr with an absolute distance modulus of $14.60 \pm 0.09$.  In all cases, the comparison models included diffusion.

4.  While the age determined for M92 appears to be too large, within the error bars, it is consistent with the recent WMAP results of the age of the universe of $13.7 \pm 0.2$ Gyr \citep{spergel}.  Accounting for the error bars, it is also consistent within $1-\sigma$ with the mean age of the oldest globular clusters: $12.6 \pm 1.2$ Gyr, based on the luminosity of the turn-off in \citet{krausschaboyer}.

\acknowledgments
Based on the observations made with the NASA/ESA Hubble Space Telescope, obtained from the Data Archive at the Space Telescope Science Institute, which is operated by the Association of Universities for Research in Astronomy, Inc., under NASA contract NAS 5-26555.  These observations are associated with program GO-9453.

\appendix
\section{The Luminosity Functions}

\begin{deluxetable}{lrrrrrrrrrrrr}
\tablecaption{The Ground-Based Luminosity Functions}
\tablewidth{0pt}
\tablecolumns{13}
\tablehead{
\colhead{Mag} &
\colhead{N$_{\rm{B}}$} &
\colhead{$\sigma_{\rm{B}}$}  &
\colhead{comp$_{\rm{B}}$} &
\colhead{log(N$_{\rm{B}}$)} &
\colhead{N$_{\rm{V}}$} &
\colhead{$\sigma_{\rm{V}}$}  &
\colhead{comp$_{\rm{V}}$} &
\colhead{log(N$_{\rm{V}}$)} &
\colhead{N$_{\rm{I}}$} &
\colhead{$\sigma_{\rm{I}}$}  &
\colhead{comp$_{\rm{I}}$} &
\colhead{log(N$_{\rm{I}}$)}  \\
}

\startdata
10.9 & \nodata & \nodata & \nodata & \nodata & \nodata & \nodata & \nodata & \nodata & 1.50 & 1.50 & 0.67 & 0.18 \\
11.3 & \nodata & \nodata & \nodata & \nodata & \nodata & \nodata & \nodata & \nodata & 2.71 & 1.93 & 0.74 & 0.43 \\
11.9 & \nodata & \nodata & \nodata & \nodata & \nodata & \nodata & \nodata & \nodata & 2.17 & 1.54 & 0.92 & 0.34 \\
12.1 & \nodata & \nodata & \nodata & \nodata & \nodata & \nodata & \nodata & \nodata & 1.18 & 1.18 & 0.85 & 0.07 \\
12.3 & \nodata & \nodata & \nodata & \nodata & 1.50 & 1.50 & 0.67 & 0.18 & 3.05 & 1.77 & 0.98 & 0.48 \\
12.5 & \nodata & \nodata & \nodata & \nodata & 2.71 & 1.92 & 0.74 & 0.43 & 2.35 & 1.67 & 0.85 & 0.37 \\
12.7 & \nodata & \nodata & \nodata & \nodata & \nodata & \nodata & \nodata & \nodata & 5.58 & 2.52 & 0.90 & 0.75 \\
12.9 & \nodata & \nodata & \nodata & \nodata & \nodata & \nodata & \nodata & \nodata & 6.55 & 2.70 & 0.92 & 0.82 \\
13.1 & \nodata & \nodata & \nodata & \nodata & 2.17 & 1.54 & 0.92 & 0.34 & 13.76 & 3.90 & 0.95 & 1.14 \\
13.3 & \nodata & \nodata & \nodata & \nodata & 3.22 & 1.86 & 0.93 & 0.51 & 2.08 & 1.48 & 0.96 & 0.32 \\
13.5 & 1.50 & 1.50 & 0.67 & 0.18 & 2.29 & 1.62 & 0.87 & 0.36 & 9.28 & 3.13 & 0.97 & 0.97 \\
13.7 & 2.71 & 1.92 & 0.74 & 0.43 & 7.94 & 3.00 & 0.88 & 0.90 & 12.21 & 3.58 & 0.98 & 1.09 \\
13.9 & \nodata & \nodata & \nodata & \nodata & 7.64 & 2.89 & 0.92 & 0.88 & 12.27 & 3.60 & 0.98 & 1.09 \\
14.1 & 2.17 & 1.54 & 0.92 & 0.34 & 10.34 & 3.27 & 0.97 & 1.02 & 5.14 & 2.31 & 0.97 & 0.71 \\
14.3 & 5.51 & 2.48 & 0.91 & 0.74 & 4.15 & 2.08 & 0.96 & 0.62 & 10.24 & 3.28 & 0.98 & 1.01 \\
14.5 & 3.30 & 1.91 & 0.91 & 0.52 & 12.33 & 3.56 & 0.97 & 1.09 & 16.25 & 4.15 & 0.99 & 1.21 \\
14.7 & 10.19 & 3.43 & 0.88 & 1.01 & 12.29 & 3.55 & 0.98 & 1.09 & 15.27 & 4.01 & 0.98 & 1.18 \\
14.9 & 12.44 & 3.64 & 0.97 & 1.10 & 12.25 & 3.54 & 0.98 & 1.09 & 13.13 & 3.70 & 0.99 & 1.12 \\
15.1 & 6.21 & 2.55 & 0.97 & 0.79 & 9.18 & 3.06 & 0.98 & 0.96 & 28.36 & 5.53 & 0.99 & 1.45 \\
15.3 & 11.32 & 3.45 & 0.97 & 1.05 & 13.20 & 3.66 & 0.99 & 1.12 & 24.38 & 5.11 & 0.98 & 1.39 \\
15.5 & 14.29 & 3.87 & 0.98 & 1.16 & 19.36 & 4.44 & 0.98 & 1.29 & 37.49 & 6.39 & 0.99 & 1.57 \\
15.7 & 10.20 & 3.26 & 0.98 & 1.01 & 14.13 & 3.78 & 0.99 & 1.15 & 31.32 & 5.81 & 0.99 & 1.50 \\
15.9 & 14.26 & 3.86 & 0.98 & 1.15 & 26.35 & 5.17 & 0.99 & 1.42 & 34.28 & 6.07 & 0.99 & 1.54 \\
16.1 & 17.34 & 4.27 & 0.98 & 1.24 & 21.30 & 4.65 & 0.99 & 1.33 & 60.50 & 8.25 & 0.99 & 1.78 \\
16.3 & 18.23 & 4.36 & 0.99 & 1.26 & 37.55 & 6.17 & 0.99 & 1.58 & 57.51 & 8.07 & 0.97 & 1.76 \\
16.5 & 18.16 & 4.35 & 0.99 & 1.26 & 38.38 & 6.23 & 0.99 & 1.58 & 67.29 & 8.76 & 0.98 & 1.83 \\
16.7 & 26.45 & 5.30 & 0.98 & 1.42 & 33.26 & 5.79 & 0.99 & 1.52 & 82.66 & 9.80 & 0.98 & 1.92 \\
16.9 & 37.52 & 6.36 & 0.99 & 1.57 & 61.50 & 7.88 & 0.99 & 1.79 & 54.99 & 7.82 & 0.98 & 1.74 \\
17.1 & 37.40 & 6.33 & 0.99 & 1.57 & 62.69 & 8.03 & 0.97 & 1.80 & 90.68 & 10.27 & 0.98 & 1.96 \\
17.3 & 37.32 & 6.32 & 0.99 & 1.57 & 71.24 & 8.51 & 0.98 & 1.85 & 137.97 & 13.08 & 0.97 & 2.14 \\
17.5 & 64.54 & 8.45 & 0.99 & 1.81 & 79.70 & 9.02 & 0.98 & 1.90 & 214.99 & 17.26 & 0.94 & 2.33 \\
17.7 & 64.67 & 8.54 & 0.97 & 1.81 & 78.44 & 8.94 & 0.98 & 1.90 & 317.48 & 21.87 & 0.93 & 2.50 \\
17.9 & 77.49 & 9.38 & 0.98 & 1.89 & 173.31 & 13.29 & 0.98 & 2.24 & 459.35 & 27.87 & 0.94 & 2.66 \\
18.1 & 84.71 & 9.85 & 0.98 & 1.93 & 277.04 & 17.22 & 0.94 & 2.44 & 619.16 & 33.68 & 0.95 & 2.79 \\
18.3 & 154.92 & 13.87 & 0.98 & 2.19 & 377.53 & 20.07 & 0.94 & 2.58 & 776.17 & 39.77 & 0.94 & 2.89 \\
18.5 & 323.17 & 22.10 & 0.95 & 2.51 & 519.89 & 23.53 & 0.94 & 2.72 & 956.42 & 46.45 & 0.86 & 2.98 \\
18.7 & 402.21 & 25.40 & 0.94 & 2.60 & 624.60 & 25.63 & 0.95 & 2.80 & 1038.20 & 49.25 & 0.87 & 3.02 \\
18.9 & 511.42 & 29.99 & 0.94 & 2.71 & 777.38 & 28.71 & 0.94 & 2.89 & 1331.19 & 59.16 & 0.86 & 3.12 \\
19.1 & 641.40 & 34.89 & 0.95 & 2.81 & 883.56 & 32.18 & 0.85 & 2.95 & 1501.23 & 64.31 & 0.87 & 3.18 \\
19.3 & 733.42 & 38.38 & 0.94 & 2.87 & 993.66 & 33.96 & 0.86 & 3.00 & 1754.11 & 72.07 & 0.84 & 3.24 \\
19.5 & 835.87 & 42.97 & 0.86 & 2.92 & 1179.65 & 37.06 & 0.86 & 3.07 & 2263.92 & 88.63 & 0.75 & 3.36 \\
19.7 & 893.22 & 45.04 & 0.86 & 2.95 & 1329.35 & 39.27 & 0.86 & 3.12 & 2366.36 & 89.34 & 0.76 & 3.37 \\
19.9 & 1070.88 & 52.23 & 0.86 & 3.03 & 1424.55 & 40.10 & 0.89 & 3.15 & 2551.47 & 93.03 & 0.77 & 3.41 \\
20.1 & 1162.87 & 55.35 & 0.87 & 3.07 & 1813.34 & 49.21 & 0.75 & 3.26 & 2754.03 & 99.54 & 0.73 & 3.44 \\
20.3 & 1268.80 & 58.27 & 0.88 & 3.10 & 2029.10 & 51.77 & 0.76 & 3.31 & 3147.71 & 111.43 & 0.64 & 3.50 \\
20.5 & 1500.95 & 67.81 & 0.80 & 3.18 & 1982.14 & 51.20 & 0.76 & 3.30 & 2631.77 & 98.29 & 0.68 & 3.42 \\
20.7 & 1654.86 & 74.35 & 0.75 & 3.22 & 2025.72 & 50.90 & 0.78 & 3.31 & \nodata & \nodata & \nodata & \nodata \\
20.9 & 1766.34 & 76.57 & 0.76 & 3.25 & 2036.92 & 50.45 & 0.80 & 3.31 & \nodata & \nodata & \nodata & \nodata \\
21.1 & 1732.39 & 75.63 & 0.76 & 3.24 & 2645.26 & 65.72 & 0.61 & 3.42 & \nodata & \nodata & \nodata & \nodata \\
21.3 & 1757.00 & 75.71 & 0.78 & 3.25 & 2432.63 & 61.47 & 0.64 & 3.39 & \nodata & \nodata & \nodata & \nodata \\
21.5 & 1713.49 & 74.52 & 0.80 & 3.23 & \nodata & \nodata & \nodata & \nodata & \nodata & \nodata & \nodata & \nodata \\
21.7 & 2111.74 & 91.10 & 0.65 & 3.33 & \nodata & \nodata & \nodata & \nodata & \nodata & \nodata & \nodata & \nodata \\
21.9 & 2116.93 & 93.42 & 0.62 & 3.33 & \nodata & \nodata & \nodata & \nodata & \nodata & \nodata & \nodata & \nodata \\
22.1 & 1871.44 & 86.28 & 0.66 & 3.27 & \nodata & \nodata & \nodata & \nodata & \nodata & \nodata & \nodata & \nodata \\
\enddata
\end{deluxetable}

\begin{deluxetable}{lrrrrrrrr}
\tablecaption{The HST Luminosity Functions}
\tablewidth{0pt}
\tablecolumns{13}
\tablehead{
\colhead{Mag} &
\colhead{N$_{\rm{V}}$} &
\colhead{$\sigma_{\rm{V}}$}  &
\colhead{comp$_{\rm{V}}$} &
\colhead{log(N$_{\rm{V}}$)} &
\colhead{N$_{\rm{I}}$} &
\colhead{$\sigma_{\rm{I}}$}  &
\colhead{comp$_{\rm{I}}$} &
\colhead{log(N$_{\rm{I}}$)}  \\
}
\startdata
10.7 & \nodata & \nodata & \nodata & \nodata & 1.00 & 1.00 & 1.00 & 0.00 \\
10.9 & \nodata & \nodata & \nodata & \nodata & 2.43 & 1.72 & 0.82 & 0.39 \\
11.1 & \nodata & \nodata & \nodata & \nodata & 1.97 & 1.97 & 0.51 & 0.29 \\
11.3 & \nodata & \nodata & \nodata & \nodata & 4.72 & 2.73 & 0.64 & 0.67 \\
11.5 & \nodata & \nodata & \nodata & \nodata & \nodata & \nodata & \nodata & \nodata \\
11.7 & \nodata & \nodata & \nodata & \nodata & 1.08 & 0.76 & 1.86 & 0.03 \\
11.9 & \nodata & \nodata & \nodata & \nodata & 2.54 & 1.47 & 1.18 & 0.40 \\
12.1 & 1.00 & 1.00 & 1.00 & 0.00 & \nodata & \nodata & \nodata & \nodata \\
12.3 & 2.43 & 1.72 & 0.82 & 0.39 & 3.13 & 2.22 & 0.64 & 0.50 \\
12.5 & 5.31 & 3.07 & 0.57 & 0.73 & 9.28 & 3.80 & 0.65 & 0.97 \\
12.7 & 1.38 & 0.80 & 2.17 & 0.14 & 3.07 & 2.17 & 0.65 & 0.49 \\
12.9 & 1.08 & 1.08 & 0.93 & 0.03 & 8.42 & 3.77 & 0.59 & 0.93 \\
13.1 & 1.48 & 1.04 & 1.36 & 0.17 & 7.21 & 2.95 & 0.83 & 0.86 \\
13.3 & 1.06 & 1.06 & 0.94 & 0.03 & 9.22 & 3.27 & 0.87 & 0.96 \\
13.5 & 4.75 & 2.75 & 0.63 & 0.68 & 16.22 & 4.36 & 0.86 & 1.21 \\
13.7 & 10.73 & 4.06 & 0.65 & 1.03 & 25.19 & 5.55 & 0.83 & 1.40 \\
13.9 & 8.28 & 3.71 & 0.60 & 0.92 & 16.25 & 4.37 & 0.86 & 1.21 \\
14.1 & 7.09 & 2.90 & 0.85 & 0.85 & 13.37 & 3.73 & 0.97 & 1.13 \\
14.3 & 13.24 & 4.01 & 0.83 & 1.12 & 17.38 & 4.24 & 0.98 & 1.24 \\
14.5 & 20.96 & 4.97 & 0.86 & 1.32 & 24.40 & 5.03 & 0.98 & 1.39 \\
14.7 & 22.27 & 5.14 & 0.85 & 1.35 & 20.32 & 4.58 & 0.98 & 1.31 \\
14.9 & 19.91 & 4.72 & 0.90 & 1.30 & 26.29 & 5.20 & 0.99 & 1.42 \\
15.1 & 16.39 & 4.12 & 0.98 & 1.21 & 32.82 & 5.87 & 0.98 & 1.52 \\
15.3 & 17.30 & 4.22 & 0.98 & 1.24 & 41.03 & 6.58 & 0.98 & 1.61 \\
15.5 & 28.47 & 5.43 & 0.98 & 1.45 & 55.22 & 7.66 & 0.98 & 1.74 \\
15.7 & 26.36 & 5.21 & 0.99 & 1.42 & 60.17 & 7.99 & 0.98 & 1.78 \\
15.9 & 32.73 & 5.84 & 0.98 & 1.51 & 55.84 & 7.67 & 0.99 & 1.75 \\
16.1 & 42.05 & 6.65 & 0.98 & 1.62 & 68.16 & 8.56 & 0.97 & 1.83 \\
16.3 & 55.24 & 7.64 & 0.98 & 1.74 & 87.98 & 9.79 & 0.97 & 1.94 \\
16.5 & 70.36 & 8.64 & 0.98 & 1.85 & 91.69 & 9.98 & 0.97 & 1.96 \\
16.7 & 56.03 & 7.67 & 0.98 & 1.75 & 108.80 & 10.83 & 0.98 & 2.04 \\
16.9 & 73.24 & 8.86 & 0.97 & 1.86 & 137.57 & 12.27 & 0.98 & 2.14 \\
17.1 & 87.13 & 9.73 & 0.96 & 1.94 & 205.70 & 15.45 & 0.95 & 2.31 \\
17.3 & 100.43 & 10.40 & 0.98 & 2.00 & 270.88 & 18.11 & 0.94 & 2.43 \\
17.5 & 144.85 & 12.59 & 0.98 & 2.16 & 410.81 & 22.74 & 0.96 & 2.61 \\
17.7 & 231.52 & 16.38 & 0.96 & 2.36 & 537.74 & 26.64 & 0.96 & 2.73 \\
17.9 & 355.63 & 21.01 & 0.95 & 2.55 & 757.05 & 32.76 & 0.96 & 2.88 \\
18.1 & 561.05 & 27.33 & 0.95 & 2.75 & 997.94 & 39.43 & 0.93 & 3.00 \\
18.3 & 697.94 & 31.22 & 0.96 & 2.84 & 1276.49 & 46.31 & 0.93 & 3.11 \\
18.5 & 864.78 & 35.83 & 0.95 & 2.94 & 1401.20 & 48.64 & 0.94 & 3.15 \\
18.7 & 1080.69 & 41.78 & 0.93 & 3.03 & 1561.22 & 52.28 & 0.94 & 3.19 \\
18.9 & 1279.05 & 46.23 & 0.93 & 3.11 & 1806.69 & 57.45 & 0.94 & 3.26 \\
19.1 & 1332.30 & 47.48 & 0.93 & 3.12 & 2182.04 & 66.22 & 0.88 & 3.34 \\
19.3 & 1486.76 & 51.28 & 0.94 & 3.17 & 2444.68 & 71.29 & 0.89 & 3.39 \\
19.5 & 1683.37 & 55.83 & 0.92 & 3.23 & 2823.79 & 78.90 & 0.89 & 3.45 \\
19.7 & 2021.90 & 64.00 & 0.89 & 3.31 & 2941.16 & 80.48 & 0.89 & 3.47 \\
19.9 & 2137.29 & 66.42 & 0.88 & 3.33 & 3020.60 & 81.24 & 0.91 & 3.48 \\
20.1 & 2419.77 & 72.77 & 0.89 & 3.38 & 3399.52 & 90.58 & 0.81 & 3.53 \\
20.3 & 2502.94 & 74.27 & 0.89 & 3.40 & 3081.61 & 83.45 & 0.84 & 3.49 \\
20.5 & 2475.90 & 73.40 & 0.90 & 3.39 & 2933.70 & 79.82 & 0.84 & 3.47 \\
20.7 & 2496.91 & 74.58 & 0.87 & 3.40 & 2395.66 & 68.53 & 0.86 & 3.38 \\
20.9 & 2730.19 & 80.55 & 0.82 & 3.44 & 2121.00 & 62.69 & 0.86 & 3.33 \\
21.1 & 2572.96 & 77.24 & 0.83 & 3.41 & 2191.36 & 66.77 & 0.77 & 3.34 \\
\enddata
\end{deluxetable}

\begin{deluxetable}{lrrrrrrrr}
\tablecaption{The Combined Luminosity Functions}
\tablewidth{0pt}
\tablecolumns{13}
\tablehead{
\colhead{Mag} &
\colhead{N$_{\rm{V}}$} &
\colhead{$\sigma_{\rm{V}}$}  &
\colhead{comp$_{\rm{V}}$} &
\colhead{log(N$_{\rm{V}}$)} &
\colhead{N$_{\rm{I}}$} &
\colhead{$\sigma_{\rm{I}}$}  &
\colhead{comp$_{\rm{I}}$} &
\colhead{log(N$_{\rm{I}}$)}  \\
}
\startdata
10.9 & \nodata & \nodata & \nodata & \nodata & 1.50 & 1.50 & 0.67 & 0.18 \\
11.1 & \nodata & \nodata & \nodata & \nodata & 9.51 & 5.49 & 0.32 & 0.98 \\
11.3 & \nodata & \nodata & \nodata & \nodata & 12.23 & 6.12 & 0.33 & 1.09 \\
11.5 & \nodata & \nodata & \nodata & \nodata & \nodata & \nodata & \nodata & \nodata \\
11.7 & \nodata & \nodata & \nodata & \nodata & 2.49 & 2.49 & 0.40 & 0.40 \\
11.9 & \nodata & \nodata & \nodata & \nodata & 2.17 & 1.54 & 0.92 & 0.34 \\
12.1 & 8.05 & 8.05 & 0.12 & 0.91 & 1.18 & 1.18 & 0.85 & 0.07 \\
12.3 & 7.28 & 4.21 & 0.41 & 0.86 & 10.19 & 4.56 & 0.49 & 1.01 \\
12.5 & 7.91 & 3.96 & 0.51 & 0.90 & 11.86 & 4.84 & 0.51 & 1.07 \\
12.7 & \nodata & \nodata & \nodata & \nodata & 5.58 & 2.50 & 0.90 & 0.75 \\
12.9 & 2.49 & 2.49 & 0.40 & 0.40 & 16.84 & 5.61 & 0.53 & 1.23 \\
13.1 & 2.17 & 1.54 & 0.92 & 0.34 & 20.97 & 4.81 & 0.91 & 1.32 \\
13.3 & 3.22 & 1.86 & 0.93 & 0.51 & 10.18 & 3.39 & 0.88 & 1.01 \\
13.5 & 11.05 & 4.94 & 0.45 & 1.04 & 25.50 & 5.32 & 0.90 & 1.41 \\
13.7 & 15.83 & 5.01 & 0.63 & 1.20 & 35.45 & 6.58 & 0.82 & 1.55 \\
13.9 & 17.79 & 5.63 & 0.56 & 1.25 & 27.41 & 5.48 & 0.91 & 1.44 \\
14.1 & 17.43 & 4.36 & 0.92 & 1.24 & 16.49 & 4.12 & 0.97 & 1.22 \\
14.3 & 16.27 & 4.35 & 0.86 & 1.21 & 26.62 & 5.22 & 0.98 & 1.43 \\
14.5 & 34.18 & 6.35 & 0.85 & 1.53 & 38.66 & 6.27 & 0.98 & 1.59 \\
14.7 & 31.71 & 5.99 & 0.88 & 1.50 & 35.58 & 6.01 & 0.98 & 1.55 \\
14.9 & 30.05 & 5.68 & 0.93 & 1.48 & 34.38 & 5.90 & 0.99 & 1.54 \\
15.1 & 23.55 & 4.91 & 0.98 & 1.37 & 59.16 & 7.77 & 0.98 & 1.77 \\
15.3 & 28.51 & 5.39 & 0.98 & 1.45 & 61.34 & 7.92 & 0.98 & 1.79 \\
15.5 & 47.83 & 6.98 & 0.98 & 1.68 & 85.56 & 9.34 & 0.98 & 1.93 \\
15.7 & 37.47 & 6.16 & 0.99 & 1.57 & 86.16 & 9.40 & 0.98 & 1.94 \\
15.9 & 56.06 & 7.56 & 0.98 & 1.75 & 88.11 & 9.45 & 0.99 & 1.95 \\
16.1 & 60.31 & 7.85 & 0.98 & 1.78 & 119.41 & 11.04 & 0.98 & 2.08 \\
16.3 & 82.59 & 9.18 & 0.98 & 1.92 & 137.33 & 11.91 & 0.97 & 2.14 \\
16.5 & 104.43 & 10.34 & 0.98 & 2.02 & 148.79 & 12.36 & 0.98 & 2.17 \\
16.7 & 85.20 & 9.30 & 0.99 & 1.93 & 181.88 & 13.67 & 0.97 & 2.26 \\
16.9 & 126.53 & 11.36 & 0.98 & 2.10 & 186.55 & 13.79 & 0.98 & 2.27 \\
17.1 & 144.65 & 12.23 & 0.97 & 2.16 & 278.08 & 17.02 & 0.96 & 2.44 \\
17.3 & 158.51 & 12.73 & 0.98 & 2.20 & 389.58 & 20.25 & 0.95 & 2.59 \\
17.5 & 215.00 & 14.87 & 0.97 & 2.33 & 597.74 & 25.10 & 0.95 & 2.78 \\
17.7 & 295.86 & 17.49 & 0.97 & 2.47 & 798.94 & 29.13 & 0.94 & 2.90 \\
17.9 & 497.36 & 22.82 & 0.96 & 2.70 & 1154.05 & 34.91 & 0.95 & 3.06 \\
18.1 & 793.09 & 28.96 & 0.95 & 2.90 & 1506.07 & 40.25 & 0.93 & 3.18 \\
18.3 & 1009.64 & 32.71 & 0.94 & 3.00 & 1907.65 & 45.36 & 0.93 & 3.28 \\
18.5 & 1305.95 & 37.33 & 0.94 & 3.12 & 2190.85 & 49.24 & 0.90 & 3.34 \\
18.7 & 1585.66 & 41.33 & 0.93 & 3.20 & 2428.96 & 51.88 & 0.90 & 3.39 \\
18.9 & 1908.09 & 45.23 & 0.93 & 3.28 & 2916.14 & 56.89 & 0.90 & 3.46 \\
19.1 & 2057.72 & 47.83 & 0.90 & 3.31 & 3410.82 & 62.72 & 0.87 & 3.53 \\
19.3 & 2295.62 & 50.40 & 0.90 & 3.36 & 3909.42 & 67.44 & 0.86 & 3.59 \\
19.5 & 2671.86 & 54.73 & 0.89 & 3.43 & 4688.35 & 75.42 & 0.82 & 3.67 \\
19.7 & 3100.09 & 59.62 & 0.87 & 3.49 & 4890.62 & 76.79 & 0.83 & 3.69 \\
19.9 & 3331.05 & 61.73 & 0.87 & 3.52 & 5108.86 & 77.97 & 0.84 & 3.71 \\
20.1 & 3887.75 & 68.44 & 0.83 & 3.59 & 5669.78 & 85.82 & 0.77 & 3.75 \\
20.3 & 4169.43 & 70.99 & 0.83 & 3.62 & 5718.65 & 88.78 & 0.73 & 3.76 \\
20.5 & 4081.43 & 69.98 & 0.83 & 3.61 & \nodata & \nodata & \nodata & \nodata \\
20.7 & 4158.33 & 70.97 & 0.83 & 3.62 & \nodata & \nodata & \nodata & \nodata \\
20.9 & 4402.63 & 74.05 & 0.80 & 3.64 & \nodata & \nodata & \nodata & \nodata \\
21.1 & 4744.25 & 81.59 & 0.71 & 3.68 & \nodata & \nodata & \nodata & \nodata \\
\enddata
\end{deluxetable}

\end{document}